\documentclass[pre,twocolumn,amssymb]{revtex4}

\usepackage{float}
\usepackage{graphicx}
\usepackage{color}
\usepackage{epsfig}
\usepackage{latexsym}
\usepackage{bm}
\usepackage{ulem}
\usepackage{color}

\begin{document}

\renewcommand{\ni}{{\noindent}}
\newcommand{\dprime}{{\prime\prime}}
\newcommand{\be}{\begin{equation}}
\newcommand{\ee}{\end{equation}}
\newcommand{\bea}{\begin{eqnarray}}
\newcommand{\eea}{\end{eqnarray}}
\newcommand{\nn}{\nonumber}
\newcommand{\bk}{{\bf k}}
\newcommand{\bQ}{{\bf Q}}
\newcommand{\q}{{\bf q}}
\newcommand{\s}{{\bf s}}
\newcommand{\bN}{{\bf \nabla}}
\newcommand{\bA}{{\bf A}}
\newcommand{\bE}{{\bf E}}
\newcommand{\bj}{{\bf j}}
\newcommand{\bJ}{{\bf J}}
\newcommand{\bs}{{\bf v}_s}
\newcommand{\bn}{{\bf v}_n}
\newcommand{\bv}{{\bf v}}
\newcommand{\la}{\langle}
\newcommand{\ra}{\rangle}
\newcommand{\dg}{\dagger}
\newcommand{\br}{{\bf{r}}}
\newcommand{\brp}{{\bf{r}^\prime}}
\newcommand{\bq}{{\bf{q}}}
\newcommand{\hx}{\hat{\bf x}}
\newcommand{\hy}{\hat{\bf y}}
\newcommand{\bS}{{\bf S}}
\newcommand{\cU}{{\cal U}}
\newcommand{\cD}{{\cal D}}
\newcommand{\bR}{{\bf R}}
\newcommand{\pll}{\parallel}
\newcommand{\sumr}{\sum_{\vr}}
\newcommand{\cP}{{\cal P}}
\newcommand{\cQ}{{\cal Q}}
\newcommand{\cS}{{\cal S}}
\newcommand{\ua}{\uparrow}
\newcommand{\da}{\downarrow}
\newcommand{\red}{\textcolor {red}}
\newcommand{\blu}{\textcolor {blue}}
\newcommand{\1}{{\oldstylenums{1}}}
\newcommand{\2}{{\oldstylenums{2}}}
\newcommand{\mDelta}{\varepsilon}
\newcommand{\m}{\tilde m}
\def\lsim {\protect \raisebox{-0.75ex}[-1.5ex]{$\;\stackrel{<}{\sim}\;$}}
\def\gsim {\protect \raisebox{-0.75ex}[-1.5ex]{$\;\stackrel{>}{\sim}\;$}}
\def\lsimeq {\protect \raisebox{-0.75ex}[-1.5ex]{$\;\stackrel{<}{\simeq}\;$}}
\def\gsimeq {\protect \raisebox{-0.75ex}[-1.5ex]{$\;\stackrel{>}{\simeq}\;$}}

\title{Additivity and density fluctuations in Vicsek-like models of self-propelled particles }

\author{Subhadip Chakraborti} 
\author{Punyabrata Pradhan}

\affiliation{ Department of Theoretical Sciences, S. N. Bose National Centre for Basic Sciences, Block-JD, Sector-III, Salt Lake, Kolkata 700106, India. 
}

\begin{abstract}

\noindent{ We study coarse-grained density fluctuations in the disordered phase of the paradigmatic Vicsek-like models of self-propelled particles with alignment interactions and random self-propulsion velocities. By numerically integrating a fluctuation-response relation - the direct consequence of an additivity property, we compute logarithm of the large-deviation probabilities of the coarse-grained subsystem density, while the system is in the disordered fluid phase with vanishing macroscopic velocity. The large-deviation probabilities, computed within additivity, agree remarkably well with that obtained from direct microscopic simulations of the models. Our results provide an evidence of the existence of an equilibrium-like chemical potential, which governs the coarse-grained density fluctuations in the Vicsek-like models. Moreover, comparison of the particle-number fluctuations among several self-propelled particle systems suggests a common mechanism through which the number fluctuations arise in such systems. }

\typeout{polish abstract}

\end{abstract}

\maketitle

\section{Introduction}  

Large-scale behaviors of self-propelled particles, such as bacterial colony, fish schools, flocks of birds, swarm of insects or an assembly of photo-activated Janus particles - collectively called {\it active matters}, have drawn much attention in the past \cite{review-Cates, review-Marchetti}. Self-propelled particles usually propel themselves by converting the chemical energy, either provided internally by the particles itself or externally by the solvents, to the mechanical one, which is then dissipated in the surrounding medium. They can be point-like, rod-like or spherical in shape, and polar or apolar in nature. Due to the mechanism of self-propulsion, such a system remains inherently out of equilibrium and could eventually reach a nonequilibrium steady state, which, in an intriguing interplay between drive, dissipation and interactions, exhibit striking collective behaviors, otherwise not possible in equilibrium.

There has been considerable progress in understanding various static and dynamic properties of self-propelled particles \cite{Vicsek_PRL1995, Toner-Tu, Fily-Marchetti, Chate_PRL2004, Chate_PRE2008}; for reviews, see \cite{review-Marchetti, review-Cates}. However, several key issues, largely revolving around deriving exact hydrodynamics \cite{Bertin, Ihle} and formulating a steady-state thermodynamics \cite{Vicsek_JPhysA1997, Cates, Brady-Takatori, Speck, Marconi, Subhadip_PRE2016} for these systems, are not yet settled. Recently, there has been a surge of interest in exploring whether fluctuations in self-propelled particle systems could be characterized by an equilibrium-like intensive thermodynamic variable, such as temperature \cite{Berthier}, chemical potential \cite{Cates_PRL2013, Subhadip_PRE2016} or pressure  \cite{Kafri1}. 
In a slightly different direction, there have been several studies to understand steady-state properties of self-propelled particles by using the methods of stochastic thermodynamics \cite{Seifert_PRL2005}, where one attempts to quantify violation of time-reversibility by defining an entropy production \cite{Debashis, Cates_PRL2016, Nardini, Mandal}. Though such an approach is quite promising, the definition of entropy production may not be unique due to the coarse-graining involved in the microscopic dynamics of such systems \cite{Puglisi_entropy, Seifert_JPhysA, Marchetti_PRE2018}.

Indeed, the major goal of constructing a steady-state thermodynamics is to understand fluctuations in systems having a nonequilibrium steady state. However, the exact nature of fluctuations in active matter systems, especially that of a conserved quantity, such as  particle number, is far from being fully understood.  
Not surprisingly, a suitable statistical mechanics framework for active matters, though highly desirable, remains elusive. In this scenario, we show that a class of self-propelled particles with Vicsek-like interactions has a simple thermodynamic characterization in terms of an equilibrium-like additivity property.

According to the standard formulation of statistical mechanics for an  equilibrium system \cite{Kardar}, where detailed balance is obeyed and consequently currents vanish, the probability weights of microscopic configurations are provided by the Boltzmann-Gibbs distributions, irrespective of the details of the microscopic dynamics of the system. In other words, in equilibrium, the probabilities of the microscopic configurations are {\it a priori} known and, therefore, can be readily used to determine the macroscopic properties of the system. However, there is no such unified principle for the systems having a nonequilibrium steady state - arguably the closest counterpart to equilibrium. Indeed, as the violation of detailed balance leads to the persistent currents in the configuration space, the probability weights of the microscopic configurations in a nonequilibrium steady state cannot be described in general by the Boltzmann-Gibbs distribution. Moreover, they are usually a priori {\it not} known and have to be obtained from the microscopic dynamics of the systems. The difficulties begin precisely at this stage: In the absence of the knowledge of the steady-state probability weights, it is in principle not possible in nonequilibrium to relate the macroscopic properties of a system to the microscopic ones.
Quite encouragingly though, an equilibrium-like approach to construct a thermodynamic framework has recently emerged for characterizing a broad class of nonequilibrium systems, that do not obey detailed balance and are not described by the Boltzmann-Gibbs distributions \cite{Eyink1996, Bertin_PRL2006, Chatterjee_PRL2014, Das_PRE2016}.

The thermodynamic property of additivity implies that, on a large scale, a system can be divided into many subsystems, which are statistically almost independent, except there are certain constraints due to the conservation laws, e.g., the conservation of mass or particle number, etc.
Additivity property is well established in equilibrium systems with short-range interactions, and, in fact, can be proven rigorously from the Boltzmann-Gibbs distribution \cite{Domb-Lebowitz}. Interestingly, also a broad class of nonequilibrium mass transport processes, having nonzero spatial correlations, can possess an additivity property in their respective nonequilibrium steady states.
As demonstrated in the past \cite{Chatterjee_PRL2014, Das_PRE2016}, provided additivity holds, subsystem mass (particle number) distributions can be determined in the thermodynamic limit directly from the knowledge of {\it only} variance of subsystem mass as a function of mass density, irrespective of whether the systems are in or out of equilibrium. 
Clearly, additivity could provide a useful simplification in  characterizing fluctuations in the driven systems as, in that case, the macroscopic properties of the systems can be determined solely through the variance of some conserved variables in the systems. However, verifying additivity is not an easy task in general.
Of course, additivity can in principle be checked analytically by explicitly calculating the steady-state subsystem mass distributions. 
But, in most cases, it may not be possible to calculate the exact functional form of the variance of subsystem mass, which is then required to obtain the subsystem mass distribution analytically \cite{Chatterjee_PRL2014, Das_PRE2016}. As we demonstrate here, this particular difficulty can be bypassed by using a numerical method instead.

In this paper, we characterize, through an additivity property, particle-number or coarse-grained density fluctuations in the paradigmatic models of self-propelled particles, namely the Vicsek model and its variant.  For simplicity, we consider only the cases when the systems are in an isotropic, and homogeneous, disordered fluid phase, where macroscopic velocity vanishes in the thermodynamic limit.
We specifically ask whether particle-number fluctuations in the Vicsek-like systems can be described by an intensive thermodynamic variable, analogous to an equilibrium-like chemical potential. We answer the question in the affirmative, by using a computational scheme within additivity. We numerically integrate a fluctuation-response relation - the direct consequence of an additivity property, to obtain a nonequilibrium chemical potential and a free energy density function, which characterize the density fluctuations in the systems. By using the above two thermodynamic potentials, we compute logarithm of the probabilities of the density large-deviations, also called the large deviation functions, and compare them with that obtained from simulations, by performing a scaling analysis of the subsystem number distributions for various subsystem sizes. We find theory and simulations are in excellent agreement, providing a strong evidence of the existence of an equilibrium-like chemical potential, which governs density fluctuations in the Vicsek-like models of self-propelled particles.

The organization of the paper is as follows. In Sec. \ref{sec-additivity}, we discuss additivity property and describe the methodology adopted in this paper for numerical computation of subsystem particle-number distributions and density large deviation functions. In Sec. \ref{models}, we present our theoretical and simulation results for two model systems - the Vicsek model in Sec. \ref{model-V0} and its variant in Sec. \ref{model-V1}. In Sec. \ref{comparison}, we compare the number fluctuations in various self-propelled particle systems and that obtained from a hydrodynamic theory of Ref. \cite{Subhadip_PRE2016}. Finally, in Sec. \ref{summary}, we summarize with some concluding remarks.

\section{Additivity and subsystem particle-number distribution} 
\label{sec-additivity}

In this section, we discuss the numerical scheme, which is used to calculate the probability distribution of subsystem particle-number. Let us first elucidate additivity property in the context of the particle-number fluctuations. We consider a system, consisting of $N$ interacting particles, which are confined in a volume $V$. Importantly, the total particle-number $N$ is conserved in the system. We then divide the system in a large number $\nu=V/v$ of identical subsystems, each having a volume $v$. Provided that the subsystem size $l=v^{1/d}$, $d$ being the dimension, is much larger than the microscopic spatial correlation length in the system, we postulate an additivity property, implying the following. The joint probability distribution ${\rm Prob.}[\{n_k\}]$ of subsystem particle numbers $\{n_k\}$, with $n_k$ being particle number in $k$th subsystem, then can have a product form \cite{Eyink1996, Bertin_PRL2006, Chatterjee_PRL2014, Das_PRE2016},
\be 
{\rm Prob.}[\{n_k\}] \simeq \frac{\prod_{k=1}^{\nu} W_v(n_k)}{Z(N, V)} \delta\left( \sum_k n_k - N \right), \label{additivity1} 
\ee 
in the thermodynamic limit of $N, V \rightarrow \infty$ with the global number density $\rho=N/V$ fixed. In the above equation, $W_v(n_k)$ is an unknown weight factor which has to be determined (discussed below), $Z = \sum_{\{n_k\}} \prod_k W_v(n_k) \delta(\sum_k n_k - N)$ is the normalization constant, or the partition sum. Note that, in Eq. (\ref{additivity1}), the weight factor $W_v(n_k)$ for the $k$th subsystem depends only on the particle-number $n_k$ and volume $v$ of the $k$th subsystem only, not on the other subsystems. Therefore, the postulate of additivity is nothing but an assumption of a statistical independence, which could emerge on the macroscopic scale, even when there are finite correlations present at the  microscopic scales.

Let us denote the subsystem particle-number distribution function as $$P_v(n) \equiv {\rm Prob.}[n_k = n] = \sum_{\{n_j; j \ne k\}} {\rm Prob.}[\{n_j\}],$$ which is the probability that a subsystem, say the $k$th one, of volume $v$ has $n$ number of particles. By using the standard statistical mechanics theory of large deviations and additivity property Eq. \ref{additivity1} \cite{Kardar, Eyink1996, Bertin_PRL2006, Chatterjee_PRL2014, Das_PRE2016}, the particle-number distribution for the $k$th subsystem can be written as
\bea
P_v(n) &\simeq& \frac{W_v(n)}{Z(N,V)} \sum_{\{n_j; j \ne k\}} \prod_{j \ne k} W_v(n_j) \delta \left( \sum_{j \ne k} n_j - N + n\right)
\nonumber
\\
&=& W_v(n) \frac{Z(N-n, V-v)}{Z(N,V)}.
\eea
One can now expand $Z(N-n, V-v)$ in the leading order of $n$ \cite{Kardar} and obtain, in the thermodynamic limit,
\be 
P_v(n) \simeq \frac{1}{\cal Z} W_v(n) e^{\mu(\rho) n},
\label{Pn1}
\ee 
where 
\be \mu(\rho)= \frac{df(\rho)}{d\rho} \label{mu-f} \ee 
is a nonequilibrium chemical potential, $f(\rho) = - \lim_{V \rightarrow \infty} \ln Z(N,V)/V$ is the corresponding nonequilibrium free energy density function  and ${\cal Z}(\mu, v)=\sum_n W_v(n) \exp[\mu(\rho) n]$ is the normalization constant. In the above equations, the symbol `$\simeq$' means equality in terms  of the logarithm $\ln P_v(n)$ of the large-deviation probability, which  can be alternatively written as
\be 
\lim_{v \rightarrow \infty} \frac{\ln [{\cal Z} P_v({n} = \hat \rho v)]}{v} = - f(\hat \rho) + \mu(\rho) \hat \rho,
\label{Pn2}
\ee 
where we denote coarse-grained subsystem density as $\hat \rho =n/v$; by definition, average of coarse-grained subsystem density equals to the global density, $\langle \hat \rho \rangle = \rho$. For the details of the above analysis, we refer to Refs.  \cite{Subhadip_PRE2016, Chatterjee_PRL2014, Das_PRE2016}.

The crucial point in this analysis is that the nonequilibrium free energy density and chemical potential can be determined as a function of number density, by using Eqs. (\ref{Pn1}) and (\ref{Pn2}) and then by integrating a fluctuation-response (FR) relation,
\be
\frac{d \rho}{d \mu} = \sigma^2(\rho),
\label{FR1}
\ee 
between a nonequilibrium compressibility and number fluctuation  - the direct consequence of additivity Eq. \ref{additivity1} (or, alternatively, Eq. \ref{Pn1}). Here, the scaled variance $\sigma^2(\rho)$ of the subsystem particle-number $n$ in the subvolume $v$ is defined as
\be
\sigma^2(\rho) = \lim_{v \rightarrow \infty} \frac{(\langle n^2 \rangle - \langle n \rangle^2)}{v}.
\label{sigma}
\ee 
For the detailed derivation of Eq. \ref{FR1} from additivity property Eq. \ref{additivity1}, see Ref. \cite{Subhadip_PRE2016}.
Now the equilibrium-like thermodynamic potentials $\mu(\rho)$ and $f(\rho)$ can be readily calculated by integrating Eqs. (\ref{mu-f}) and (\ref{FR1}) with respect to density $\rho$ and can be expressed in the integral form,
\be 
\mu(\rho) = \int \frac{1}{\sigma^2(\rho)} d\rho + c_1,
\label{mu_rho}
\ee 
and 
\be
f(\rho) = \int \mu(\rho) d\rho + c_2,
\label{f_rho}
\ee 
where $c_1$ and $c_2$ are two arbitrary constants of integrations. In this paper, we calculate chemical potential $\mu(\rho) = \int_{\rho_0}^{\rho} 1/\sigma^2(\rho) d\rho$ and free energy density $f(\rho)=\int_{\rho_0}^{\rho} \mu(\rho) d\rho$ by integrating from a reference density $\rho_0$. Finally, the large-deviation probability $ {\cal P}_v(\hat \rho)  \equiv P_v(n=v \hat \rho)$ of subsystem density $\hat \rho = n/v$, in the limit of $v$ large, can be written as \cite{Subhadip_PRE2016}
\be 
{\cal P}_v(\hat \rho) \simeq \frac{\exp[-v h(\hat \rho)]}{{\cal Z}(\mu, v)},
\label{P_rho}
\ee
or, equivalently, the subsystem number distribution is given by
\be 
P_v(n) = \frac{\exp[-vh(n/v)]}{{\cal Z}(\mu,v)},
\label{Pn3}
\ee
where the large-deviation function, or the `rate-function' \cite{Touchette}, 
\be
-h(\hat \rho) \equiv  \lim_{v \rightarrow \infty} \frac{ \ln[{\cal Z}(\mu, v) {\cal P}_v(\hat \rho)]}{v} = - f(\hat \rho) + \mu(\rho) \hat \rho,
\label{h_rho}
\ee
with $\mu(\rho)$ being chemical potential of the system at global density $\rho$ and ${\cal Z}(\mu, v) = \sum_{\hat \rho} \exp[-v h(\hat \rho)]$ being the normalization constant. Indeed, Eqs. (\ref{P_rho}), (\ref{Pn3}) and (\ref{h_rho}) can be understood from that the Laplace transform of the weight factor $W_v(n)$ is related to the Legendre transform of the nonequilibrium free energy density function - an immediate consequence of additivity Eq. (\ref{additivity1}) \cite{Subhadip_PRE2016}. Note that, through Eqs. (\ref{mu_rho}) and (\ref{f_rho}), the large-deviation probability $P_v(n)$ is thus determined solely in terms of the variance [Eq. (\ref{sigma})]. 
However, the nonequilibrium free energy function and chemical potential derived in this  paper could not be related to the mechanical pressure of the system \cite{Kafri1}, which, for a nonequilibrium system, is in principle different from the thermodynamic pressure $-h(\rho) = \mu \rho - f(\rho)$ of the system.

What remains now is to explicitly calculate the scaled variance $\sigma^2(\rho)$ of the subsystem particle number in various models through simulations, which we do next in the following sections. Moreover, by numerically integrating Eqs. (\ref{mu_rho}) and (\ref{f_rho}), we compute the subsystem particle-number distribution $P_v(n)$ and the large deviation function and then compare the number distributions with that obtained from the direct microscopic simulations.

\begin{figure}
\begin{center}
\leavevmode
\includegraphics[width=9.0cm,angle=0]{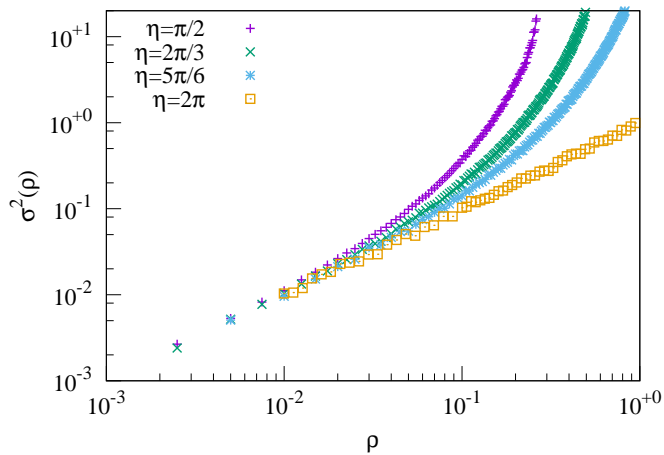}
\caption{{\it Vicsek model.} Scaled variance $\sigma^2(\rho)$ of subsystem particle number, as defined in Eq. (\ref{sigma}), is plotted as a function of number density $\rho$ for various values of noise strength $\eta=2 \pi$ (yellow squares), $5 \pi/6$ (sky-blue asterisks), $2\pi/3$ (green tilted crosses) and $\pi/2$ (violet crosses). }
\label{sigma-VM}
\end{center}
\end{figure}

\begin{figure}
\begin{center}
\leavevmode
\includegraphics[width=9.0cm,angle=0]{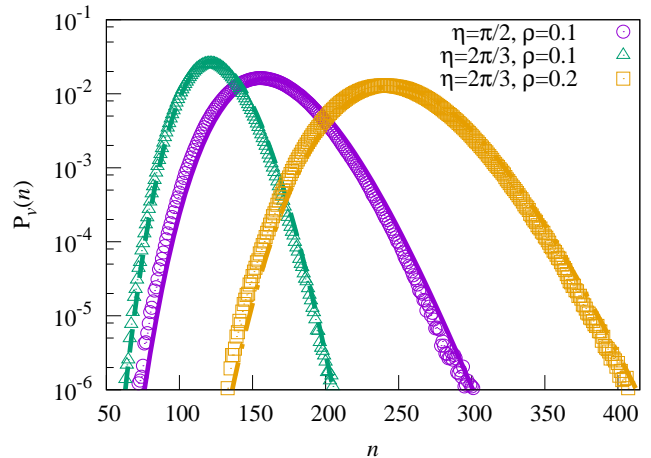}
\caption{{\it Vicsek model.} Subsystem particle-number distribution $P_v(n)$ is plotted as a function of particle number $n$ in a subsystem of volume $v$ for three sets of noise strength and density - (i) $\eta=2\pi/3$ and $\rho=0.1$ (simulation - green triangles, $v=35 \times 35$; additivity theory - green dashed line), (ii) $\eta=\pi/2$ and $\rho=0.1$ (simulation - violet circles, $v=40 \times 40$; additivity theory - violet solid line) and (iii) $\eta=2\pi/3$ and $\rho=0.2$ (simulation - yellow squares, $v=35 \times 35$; additivity theory - yellow dashed-dotted line); we take system size $L=500$. The particle-number distributions are manifestly non-Gaussian at the tails for lower noise strengths.}
\label{Pn-VM}
\end{center}
\end{figure}


\section{Models}
\label{models}

The model-systems we consider in this section consist of polar point particles on a two dimensional periodic space in continuum. At any instant of time, a particle moves with a uniform speed along a particular direction, obtained by averaging over the instantaneous directions of the neighboring particles (alignment interaction) and then by adding a noise to the instantaneous averaged direction. Depending on how the noise term is incorporated, we consider two model systems - the paradigmatic Vicsek model and its variant. In both the models, because of the polar nature of the particles, detailed balance is violated at the microscopic level. Consequently, the systems always remain out of equilibrium, eventually reaching a nonequilibrium steady state, where, unlike in equilibrium, the probability weights of the microscopic configurations cannot be described by the Boltzmann-Gibbs distribution.

\subsection{Vicsek Model}
\label{model-V0}

First we consider the Vicsek model, which consists of $N$ point particles, moving in a two dimensional periodic box of volume $V=L \times L$ in continuum \cite{Vicsek_PRL1995}. At any time $t$, the system is specified by a set of dynamical variables $\{ {\bf r}_i(t), \theta_i(t)\}$, with $i = 1, \dots, N$, where ${\bf r}_i(t)$ and $\theta_i(t)$ are the position and the self-propulsion direction of $i$th particle, respectively. The system evolves in discrete time steps according to the following dynamical rules. Each particle tries to follow its neighbors by averaging over the directions of all the neighboring particles within a circle of radius $R$. In doing so, it also makes some error of amount $\Delta \theta_i (t)$ in the direction, where $\Delta \theta _i (t)$ is a random noise variable, chosen  independently at each time step $t$ and uniformly distributed in an interval $[-\eta/2, \eta/2]$ with $\eta$ being the noise-strength. 
The equations of motion \cite{Vicsek_PRL1995} for ${\bf r}_i(t)$ and $\theta_i(t)$ of $i$th particle at a discrete time step $t$ are given by
\begin{eqnarray}
\theta _i (t+1) &=& \arctan \left[ \frac{\langle \sin \theta (t) \rangle ^R_i}{\langle \cos \theta (t) \rangle ^R_i} \right] + \Delta \theta _i (t), \\
{\bf r}_i (t+1) &=& {\bf r}_i (t) + u_0 [\cos \theta_i (t+1),\sin \theta_i (t+1)],
\end{eqnarray}
where angular bracket $\langle {\bf *} \rangle_i^R$ denotes the average of an observable over all $n^R_i$ number of neighboring particles within the radius $R=1$, i.e.,
$$
\langle \sin \theta (t) \rangle ^R_i =  \sum_{j=1}^{n^R_i} \frac{\sin \theta_j(t)}{n^R_i}~;~\langle \cos \theta (t) \rangle ^R_i = \sum_{j=1}^{n^R_i} \frac{\cos \theta_j(t)}{n^R_i},
$$
and the constant $u_0$ is the self-propulsion speed. All particles are updated in parallel; we take $u_0=0.5$ throughout. For vanishing noise strength $\eta=0$, the above dynamics becomes fully deterministic. For generic parameter values of the noise strength, the system reaches a nonequilibrium steady state at long times.

\begin{widetext}

\begin{figure}[h]
\begin{center}
\leavevmode
\hspace{-0.6cm}
\includegraphics[width=9.4cm,angle=0]{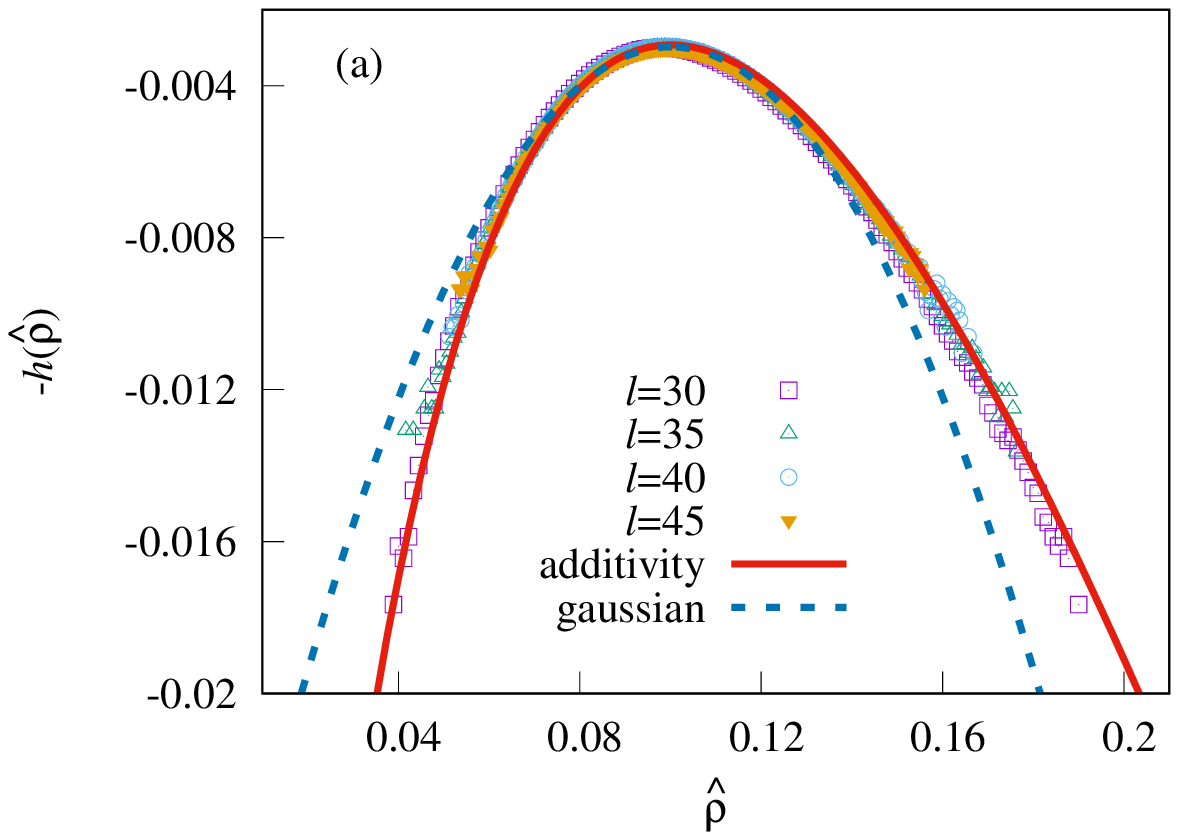}
\hspace{-0.6cm}
\includegraphics[width=9.4cm,angle=0]{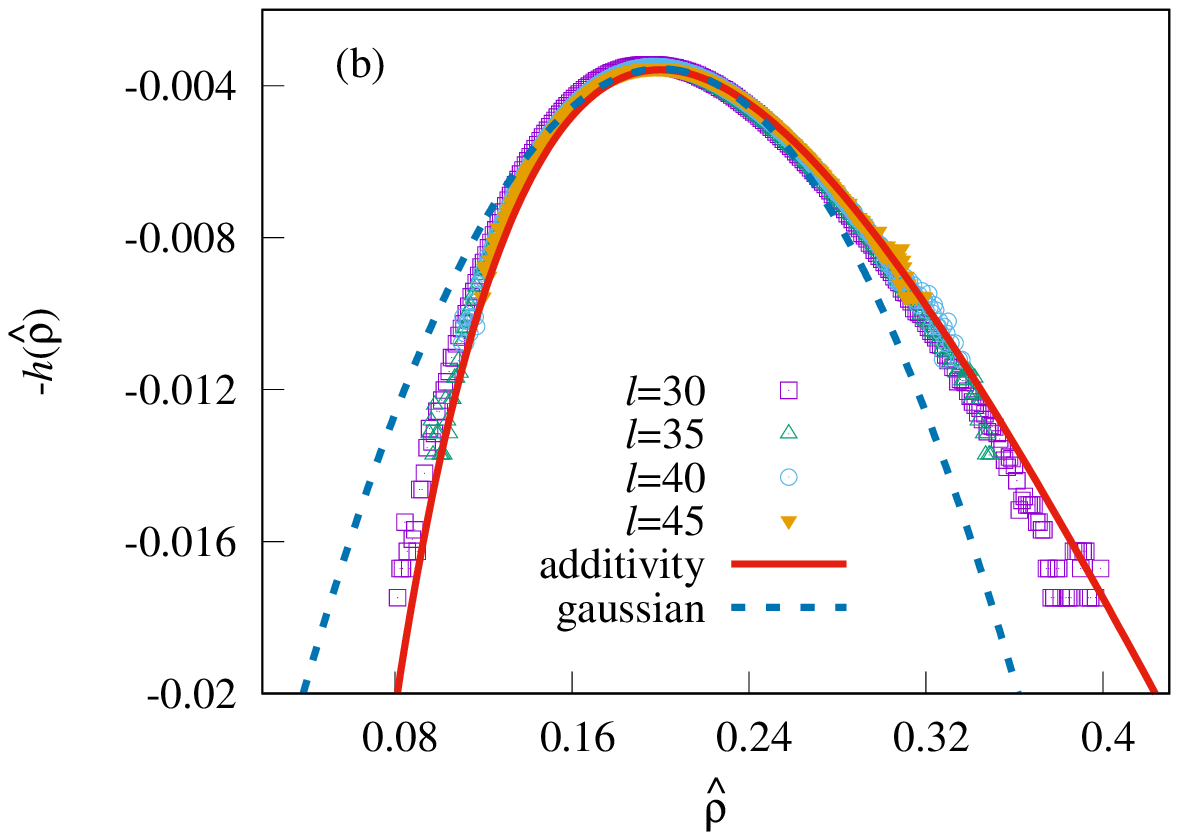}
\hspace{-0.6cm}
\includegraphics[width=9.4cm,angle=0]{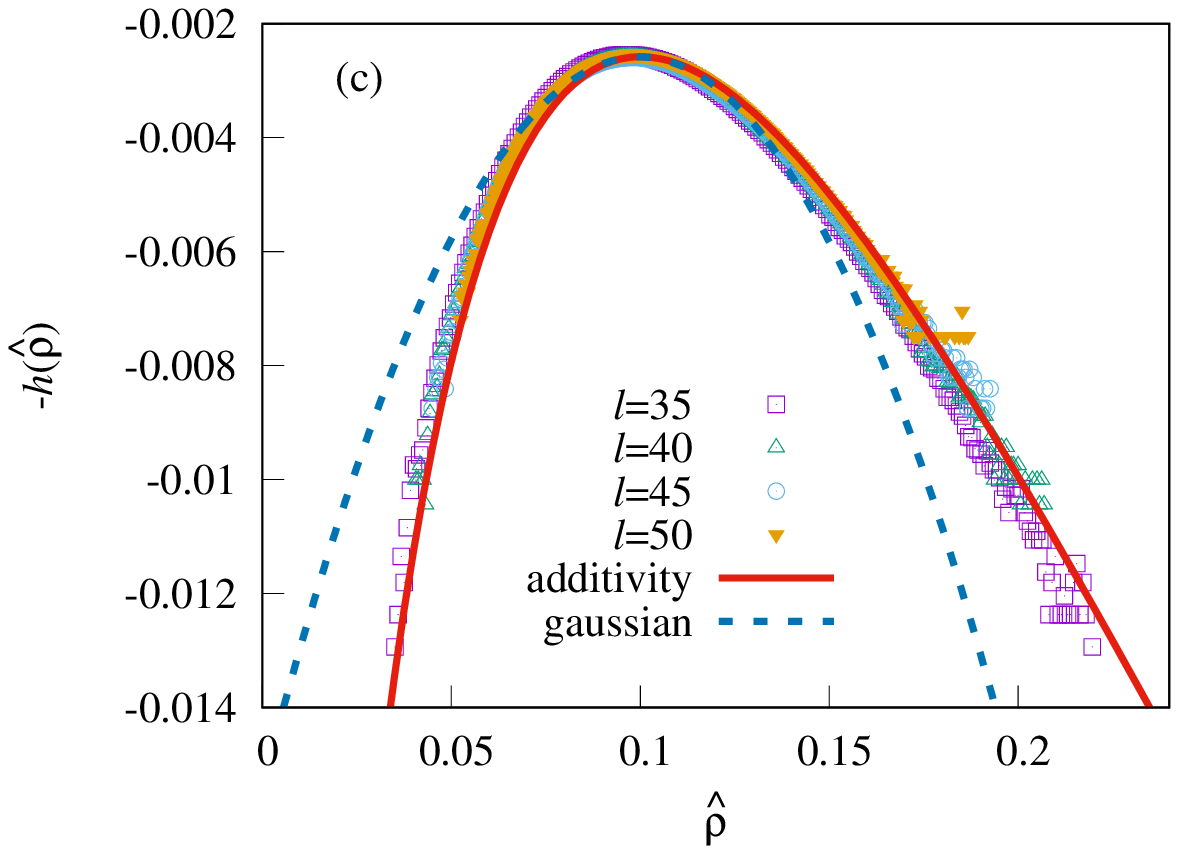}
\caption{{\it Vicsek Model.} Large deviation function $-h(\hat \rho)=\ln[{\cal Z}(\mu, v) {\cal P}_v(\hat \rho=n/v)]/v$ [Eq. (\ref{h_rho})] is plotted as a function of subsystem density $\hat \rho=n/v$, with $n$ being subsystem particle-number for various subsystem volume $v=l \times l$ with $l=30$ (violet squares), $35$ (green triangles), $40$ (sky-blue circles), $50$ (yellow inverted triangles) for three sets of noise strength and global density - (i) $\eta=2\pi/3$ and $\rho=0.1$ [panel (a)], (ii) $\eta=2\pi/3$ and $\rho=0.2$ [panel (b)] and (iii) $\eta=\pi/2$ and $\rho=0.1$ [panel (c)]. Points - simulations, red solid lines - additivity theory, sky-blue dashed lines - parabolic large deviation functions, obtained from Gaussian distributions. The non-parabolic tails, which are the consequence of non-Gaussian number distributions, are well captured by additivity theory. }
\label{Prho-VM}
\end{center}
\end{figure}

\end{widetext}

The Vicsek model has been intensively studied in the past and the nature of the phase diagram is still a debated issue in the literature \cite{Chate_PRL2004, VM-discont, SCT, Trefz}. At very large density $\rho$, or at small noise-strength $\eta$, particles are observed to move along a spontaneously selected direction, indicating a phase transition, upon tuning the number density or the noise strength, from an isotropic (rotationally symmetric) disordered phase with zero macroscopic velocity to an ordered phase with nonzero macroscopic velocity, where the rotational symmetry is spontaneously broken. The phase transition is characterized by an order parameter, which, in this case, is taken to be the magnitude of the macroscopic velocity $v_a = ({1}/{N u_0}) |\sum_{i=1}^N {\bf u}_i |$, obtained by averaging over the velocities ${\bf u}_i=\{u_0 \cos \theta_i, u_0 \sin \theta_i\}$ of all particles in the system.
For simplicity, in this paper we confine our studies to the disordered fluid phase where the macroscopic velocity vanishes ($v_a=0$). Although the macroscopic velocity is vanishingly small in the disordered phase, detailed balance in the microscopic configuration space is still violated, driving the system out of equilibrium.

As discussed in the previous section, to check whether the system possesses an additivity property, we require to characterize particle-number fluctuations on the coarse-grained level. To this end, we study subsystem particle number fluctuations as a function of number density. We consider a subsystem of volume $v=l \times l$ with the global density $\rho=N/V$ (or, equivalently, chemical potential) fixed. In our simulations throughout the paper, we keep  system sizes much larger than the microscopic length scale $R$ and the subsystem size, and the subsystem size much larger than $R$, i.e., $R \ll l \ll L$. Although the total number of particles is conserved in the system, the subsystem particle number $n$, or the coarse-grained density $\hat \rho=n/v$, fluctuates in time and is a random variable, whose statistics are of our main interest and are presented below.

Following the numerical scheme of Sec. \ref{sec-additivity}, we first calculate in direct microscopic simulations the scaled variance $\sigma^2(\rho)$ [Eq. (\ref{sigma})] of subsystem particle number. In Fig. \ref{sigma-VM}, we plot the scaled variance $\sigma^2(\rho)$ as a function of number density $\rho$ for various values of the noise strength, $\eta = 2\pi$, $5\pi/6$, $2\pi/3$ and $\pi/2$. For the  maximum possible noise strength $\eta=2\pi$, the system essentially behaves like an ideal gas of noninteracting particles and consequently the scaled variance $\sigma^2(\rho)=\rho$ increases linearly as a function of density $\rho$ \cite{Subhadip_PRE2016}. However, for lower noise strength $\eta < 2\pi$, the collective behavior sets in and the scaled variance grows quite rapidly as a function of density in a highly nonlinear fashion. The scaled variance finally diverges at a critical density, beyond which macroscopic clusters of particles are formed in the system.

Using the functional dependence of the scaled variance on density, as obtained from microscopic simulations (Fig. \ref{sigma-VM}), we numerically integrate Eqs. (\ref{mu_rho}) and (\ref{f_rho}), with respect to density $\rho$, to obtain a nonequilibrium chemical potential and free energy density, respectively. Then, using Eq. (\ref{Pn3}), we numerically compute the subsystem particle number distribution $P_v(n)$ as a function of subsystem particle number $n$ for various values of the global densities $\rho$ and the noise strengths $\eta$. We also directly calculate the number distributions $P_v(n)$ from microscopic simulations of the Vicsek model. In Fig. \ref{Pn-VM}, we plot the number distributions $P_v(n)$ as a function of $n$ for three different sets of global density and noise strength: (i) $\rho=0.1$ and $\eta=2\pi/3$ (green triangles), (ii) $\rho=0.1$ and $\eta=\pi/2$ (violet circles), and (iii) $\rho=0.2$ and $\eta=2\pi/3$ (yellow squares). We compare the number distributions from simulations with that obtained from additivity theory [Eq. (\ref{Pn3})] in Fig. \ref{Pn-VM}, where green dashed, violet solid and yellow dashed-dotted lines represent theoretically obtained number distributions for the sets (i), (ii) and (iii) of density and noise strength, respectively. As one can see, theory and simulations are in quite good agreement over several orders of magnitudes of the probabilities; notably, there is no fitting parameter involved in the theory. However, there are small deviations, which are expected though as there can be sub-leading corrections to the large-deviation functions, due to the finite size of the subsystem and the system.

Next we perform a scaling analysis to check whether the large deviation probabilities ${\cal P}_v(\hat \rho=n/v) \sim \exp[-vh(\hat \rho)]$ or, equivalently, the density large deviation functions $-h(\hat \rho)$, obtained from simulations for different subsystem sizes, converge to that obtained from additivity theory Eq. (\ref{h_rho}). In panel (a) of Fig. \ref{Prho-VM}, for a fixed global density $\rho=0.1$ [or a fixed chemical potential $\mu(\rho)$] and noise strength $\eta=2\pi/3$, we plot the large deviation functions $-h(\hat \rho)=\ln[{\cal Z}(\mu, v) {\cal P}_v(\rho)]/v$ - a suitably scaled subsystem number distribution, for various values of subsystem sizes $l=30$ (violet squares), $l=35$ (sky-blue triangles), $40$ (sky-blue diamonds) and $45$ (orange inverted triangles); we take system size $L=500$. We also compare the large deviation functions with that obtained from theory Eq. (\ref{h_rho}) (red solid lines). One could see that simulations and theory, which is without any fitting parameter,  are in quite good agreement. We repeat the above scaling analysis for the other two sets of global density and noise strength: $\rho=0.2$ and $\eta=2\pi/3$ [panel (b), Fig. \ref{Prho-VM}] and $\rho=0.1$ and $\eta=\pi/2$ [panel (c), Fig. \ref{Prho-VM}]. Moreover, to emphasize that the subsystem number distributions are not merely Gaussian distributions $\sim \exp[v(\hat \rho - \rho)^2/2 \sigma^2(\rho)]$ with mean subsystem particle number $\langle n \rangle = v\rho$ and variance $\langle n^2 \rangle - \langle n \rangle^2 = v\sigma^2$, we compare in Fig. \ref{Prho-VM} the large deviation functions, obtained from theory and simulations, with the Gaussian ones, which are simple parabolas (sky-blue dashed lines in the panels). The subsystem number distributions, beyond the central region, deviate significantly from the Gaussian distributions. Consequently, the corresponding density large-deviation functions have non-parabolic tails, which are remarkably well captured by additivity theory [Eq. (\ref{h_rho})].   

\begin{figure}[h]
\begin{center}
\leavevmode
\includegraphics[width=9.0cm,angle=0]{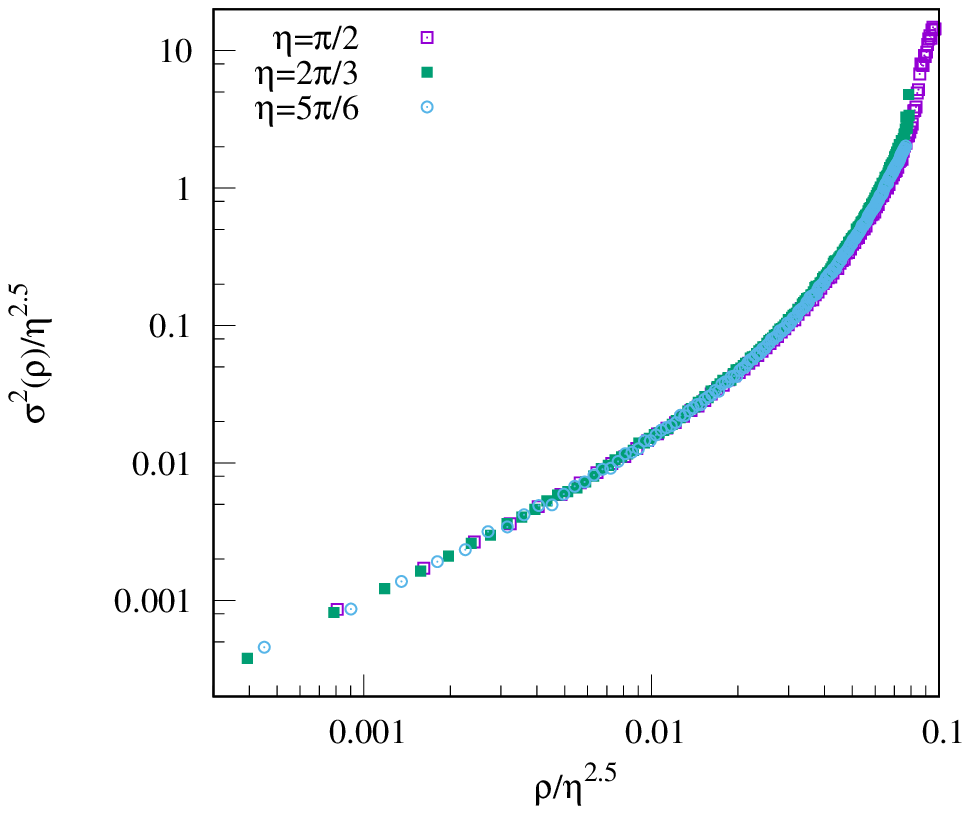}
\includegraphics[width=9.4cm,angle=0]{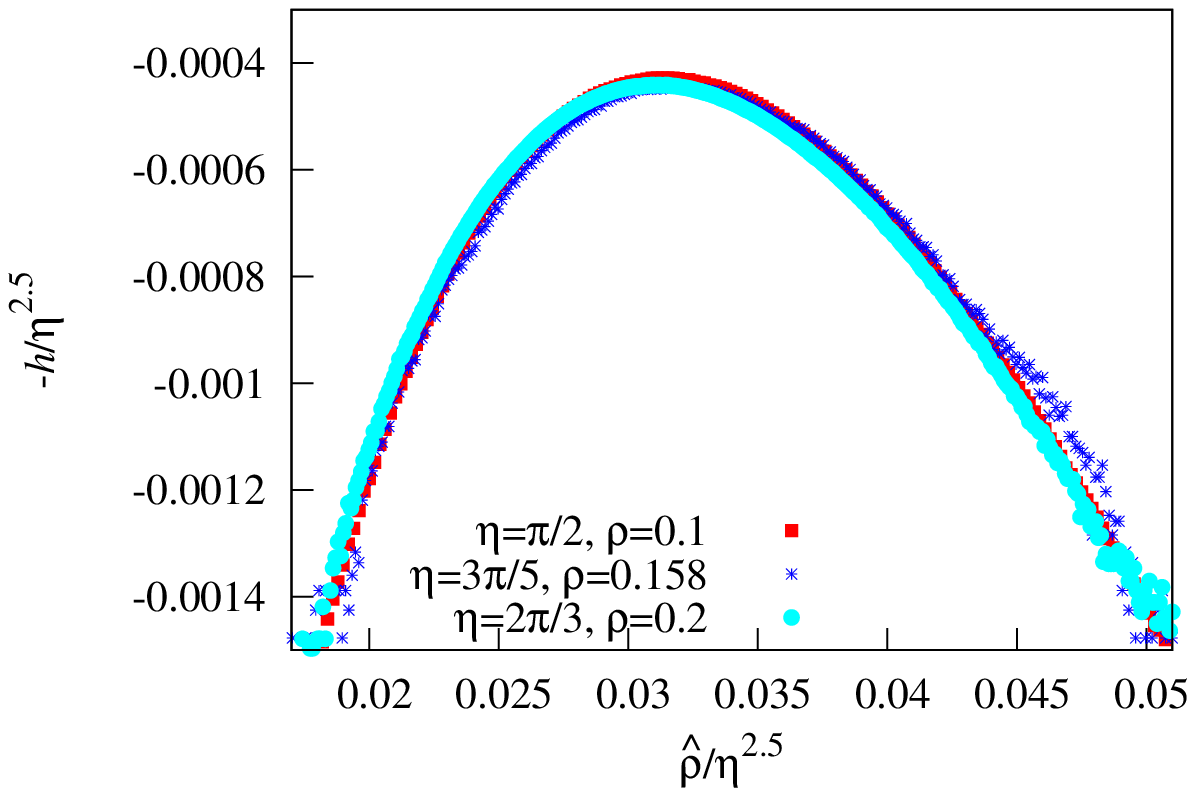}
\caption{Scaling of subsystem particle-number fluctuations in the Vicsek model in the disordered phase. Top panel: A scaling collapse is observed over several orders of magnitude of number-fluctuation and density when the rescaled variance $\sigma^2(\rho)/\eta^{\Delta}$, with $\Delta \approx 2.5$, is plotted as a function of rescaled density $\rho/\eta^{\Delta}$ for various noise strengths $\eta=\pi/2$ (violet open squares), $2\pi/3$  (green filled squares) and $5\pi/6$ (sky-blue open circles). Bottom panel: Rescaled large deviation function $h/\eta^{\Delta}$ is plotted as a function rescaled subsystem density  $\hat \rho/\eta^{\Delta}$ for three sets of noise strength $\eta$ and global density $\rho$: $\eta=\pi/2$ and $\rho=0.1$ (red squares), $\eta=3\pi/5$ and $\rho=0.158$ (blue asterisks) and $\eta=2\pi/3$ and $\rho=0.2$ (sky-blue circles) with rescaled global density $\rho/\eta^{\Delta}$ fixed at a certain value. }
\label{scaling-VM}
\end{center}
\end{figure}

From Fig. \ref{sigma-VM}, one could observe that, although the systems are in the disordered phase, the cooperative behavior, which is manifest in the significant rise in the spatial correlations and consequently the larger fluctuations in the system, increases quite rapidly with decreasing noise strength $\eta$, while one approaches the phase transition region. Moreover, the decrease in the noise strength is somewhat offset by the decrease in the number density in the systems. In other words, at a higher noise strength, one must  have a higher density to recover the same effect of cooperativity, or fluctuations, which one had at a lower noise strength and density. To quantify the above observation, we now attempt to scale the variables as $\sigma^2(\rho) \rightarrow \sigma^2(\rho)/\eta^{\Delta}$ and $\rho \rightarrow \rho/\eta^{\Delta}$. In the top panel of Fig. \ref{scaling-VM}, we plot the rescaled variance $\sigma^2(\rho)/\eta^{\Delta}$ as a function of the rescaled density $\rho/\eta^{\Delta}$ with $\Delta \approx 2.5$ for three noise strengths $\eta=\pi/2$, $2 \pi/3$ and $5 \pi/6$. Quite strikingly, we have a reasonably good scaling collapse over several decades of magnitude of fluctuations and densities. This particular scaling collapse suggests a scaling form for the variance of subsystem particle-number,
\be 
\sigma^2(\rho, \eta) = \eta^{\Delta} g\left( \frac{\rho}{\eta^{\Delta}} \right),
\label{g_scaling}
\ee 
where $g$ is a scaling function. The above scaling form in Eq. (\ref{g_scaling}) leads to the following scaling for nonequilibrium chemical potential $\mu (\rho, \eta) = \tilde \mu (\rho/\eta^{\Delta})$ and nonequilibrium free energy density $f(\rho, \eta) = \eta^{\Delta} \tilde f (\rho/\eta^{\Delta})$ where $\tilde \mu$ and $\tilde f$ are some scaling functions. Consequently the large deviation function can be written in the following form,
\bea 
h(\hat \rho, \eta) = \eta^{\Delta} \left[ \tilde f \left( \frac{\hat \rho}{\eta^{\Delta}} \right) - \tilde \mu \left( \frac{\rho}{\eta^{\Delta}} \right) \frac{\hat \rho}{\eta^{\Delta}} \right],
\eea
where $\rho$ is the global number density and $\hat \rho$ is the fluctuating coarse-grained subsystem density. In bottom panel of Fig. \ref{scaling-VM}, we plot the scaled large deviation function $-h/\eta^{\Delta}$ as a function of the scaled subsystem density $\hat \rho/\eta^{\Delta}$ with $\Delta \approx 2.5$ for three sets of global density $\rho$ and noise strength $\eta$: (i) $\rho=0.1$ and $\eta = \pi/2$, (ii)  $\rho=0.158$ and $\eta = 3\pi/5$ and (iii)  $\rho=0.2$ and $\eta = 2 \pi/3$, where we keep the ratio $\rho/\eta^{\Delta}$ fixed. A quite good scaling collapse in bottom panel of Fig. \ref{scaling-VM} confirms the scaling already observed in top panel of Fig. \ref{scaling-VM} for the subsystem particle-number fluctuations.


\subsection{Variant of the Vicsek Model}
\label{model-V1}

\begin{figure}[t]
\begin{center}
\leavevmode
\includegraphics[width=9.0cm,angle=0]{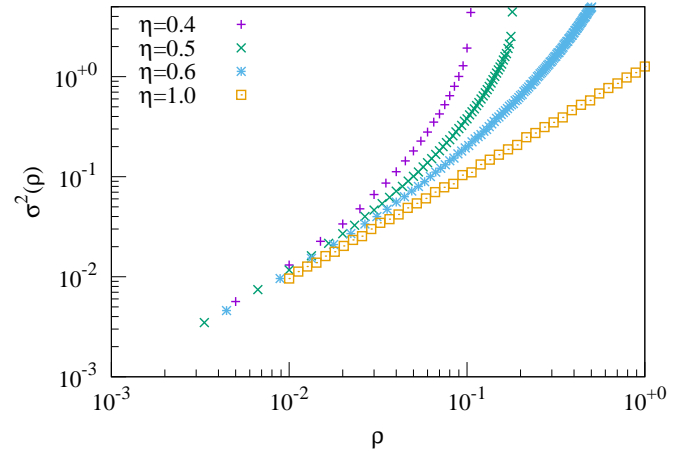}
\caption{{\it Variant of the Vicsek model.} Scaled variance $\sigma^2(\rho)$ of subsystem particle number, as defined in Eq. (\ref{sigma}), is plotted as a function of number density $\rho$ for various values of noise strength $\eta=1.0$ (yellow squares), $0.6$ (sky-blue asterisks), $0.5$ (green tilted crosses)  and $0.4$ (violet crosses).}
\label{sigma-VM1}
\end{center}
\end{figure}

\begin{figure}[t]
\begin{center}
\leavevmode
\includegraphics[width=9.0cm,angle=0]{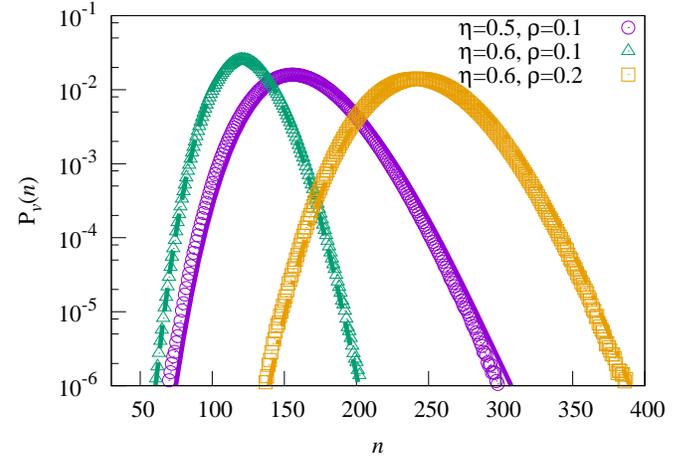}
\caption{{\it Variant of the Vicsek model.} Subsystem particle-number distribution $P_v(n)$ is plotted as a function of particle number $n$ in a subvolume $v$ for three sets of noise strength and density: (i) $\eta=0.6$ and $\rho=0.1$ (simulation - green triangles, $v=35 \times 35$; additivity theory - green dashed line), (ii) $\eta=0.5$ and $\rho=0.1$ (simulation - violet circles, $v=40 \times 40$; additivity theory - violet solid line) and (iii) $\eta=0.6$ and $\rho=0.2$ (simulation - yellow squares, $v=35 \times 35$; additivity theory - yellow dashed-dotted line); we take system size $L=500$. }
\label{Pn-VM1}
\end{center}
\end{figure}

Now we consider a variant of the Vicsek model, which consists of $N$ particles, moving in continuum  in a periodic box of volume $V=L \times L$. The variant is similar to the one introduced previously in Ref. \cite{Chate_PRL2004}, which is different from the original Vicsek model in the way the noise term, i.e., the error in estimating the direction of a particle, is incorporated in the dynamics. The error or the noise is now added separately to each of the components of the direction vector, obtained by averaging over the neighbors within a disk of radius $R=1$. In other words, the noise term in the variant constitutes a vector noise, as opposed to the scalar (angle) noise in the Vicsek model. However, apart from the noise term, the deterministic part of the dynamics is the same as in the Vicsek model. The modified equations of motion are given by
\begin{eqnarray}
\theta _i(t+1) &=& \arctan \left[ \frac{\langle \sin \theta (t) \rangle ^R_i + \eta \sin \xi _i (t)}{\langle \cos \theta (t) \rangle ^R_i + \eta \cos \xi _i (t)} \right], \\
{\bf r}_i (t+1) &=& {\bf r}_i (t) + u_0 [\cos \theta_i (t+1),\sin \theta_i (t+1)],
\end{eqnarray}
where angular bracket $\langle {\bf *} \rangle_i^R$ denotes average over all neighboring particles $j$ satisfying $|{\bf r}_j (t) - {\bf r}_i (t)| < R=1$, $\xi _i (t) \in [-\pi, \pi]$ is a uniformly distributed random angle, $\eta$ is the noise strength and $u_0=0.5$ is the self-propulsion speed of the particles. Total particle number is conserved under the above dynamics and, at long times, the system reaches a nonequilibrium steady state. As in the Vicsek model, beyond a particular value of the density and the noise strength, we observe a phase transition from an isotropic disordered fluid phase with vanishing macroscopic velocity ($v_a=0$) to an ordered phase with nonzero macroscopic velocity ($v_a \ne 0$) \cite{Chate_PRL2004, Chate_PRE2008}. The exact nature of the phase diagram for this particular noise variable is not fully understood though \cite{VM-discont}. For simplicity, we confine our studies to the disordered phase only, where the system remains homogeneous and which, due to the violation of detailed balance, is still out of equilibrium.

\begin{widetext}

\begin{figure}[t]
\begin{center}
\leavevmode
\hspace{-0.6cm}
\includegraphics[width=9.4cm,angle=0]{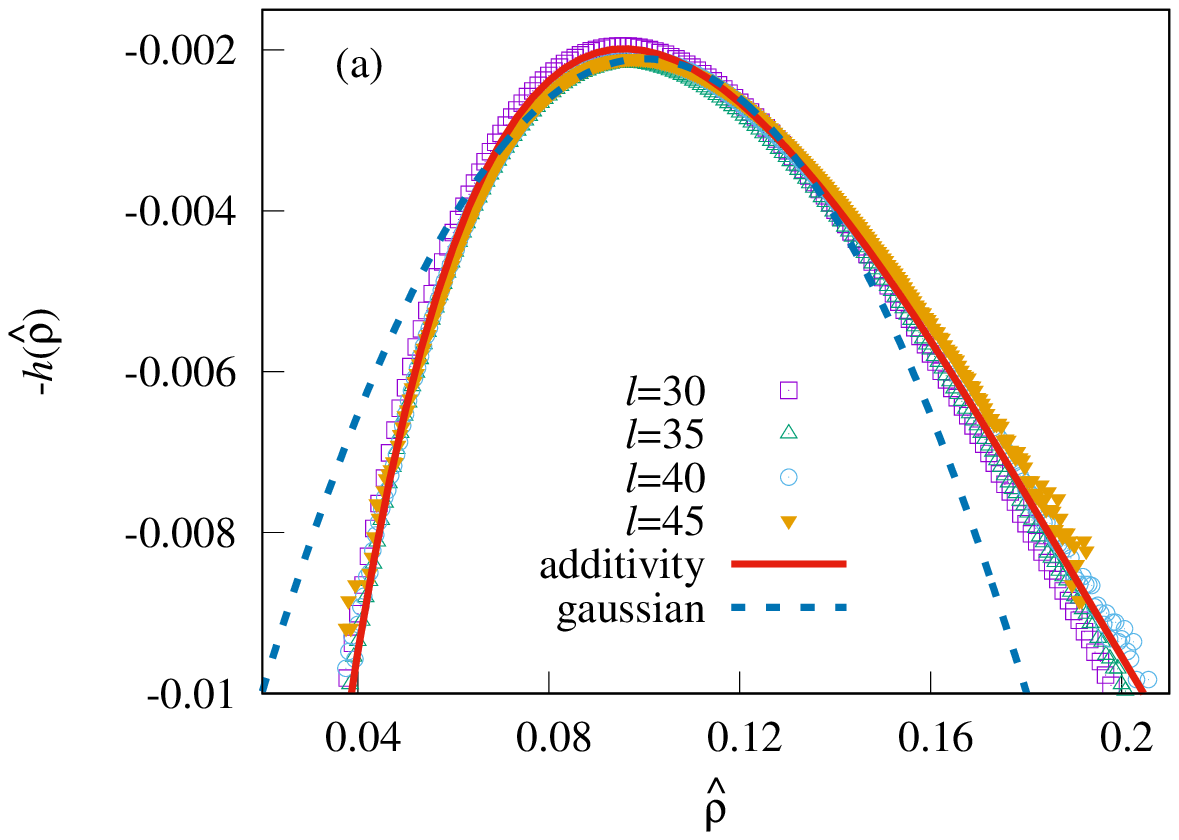}
\hspace{-0.6cm}
\includegraphics[width=9.4cm,angle=0]{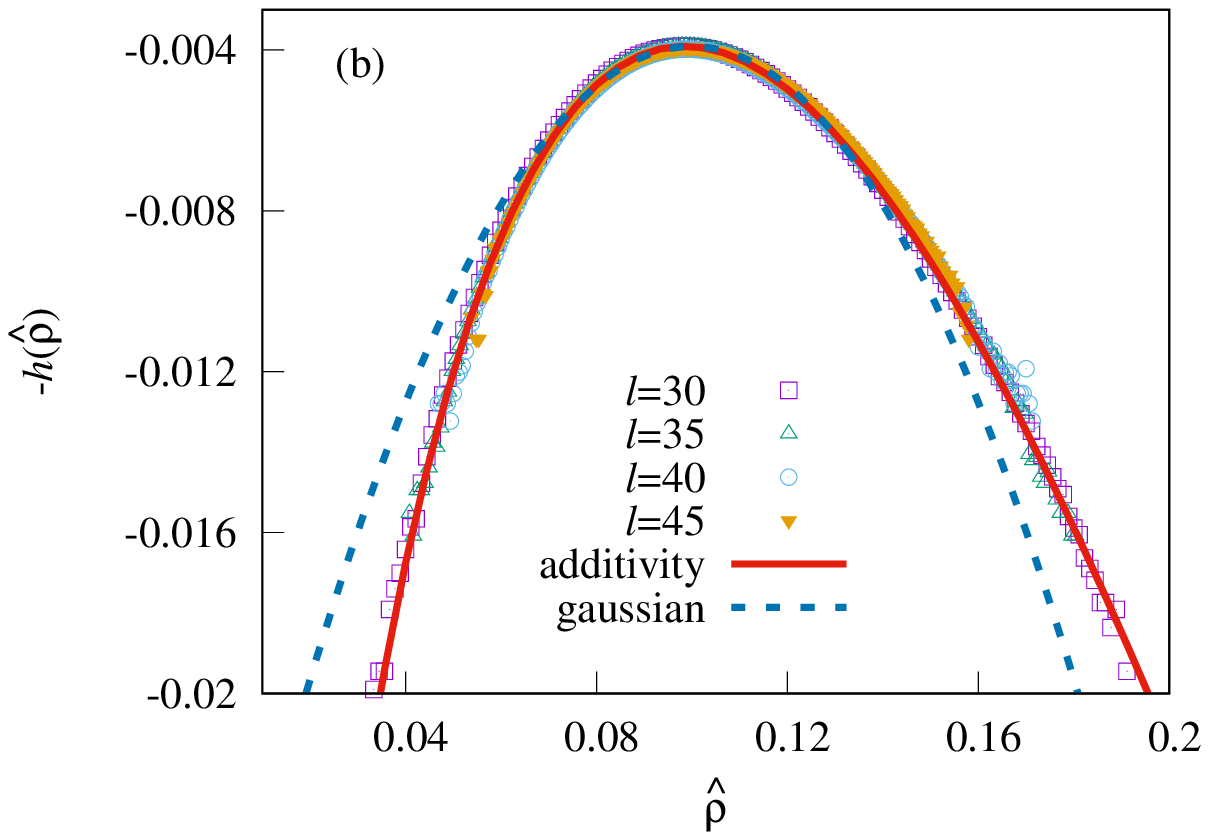}
\hspace{-0.6cm}
\includegraphics[width=9.4cm,angle=0]{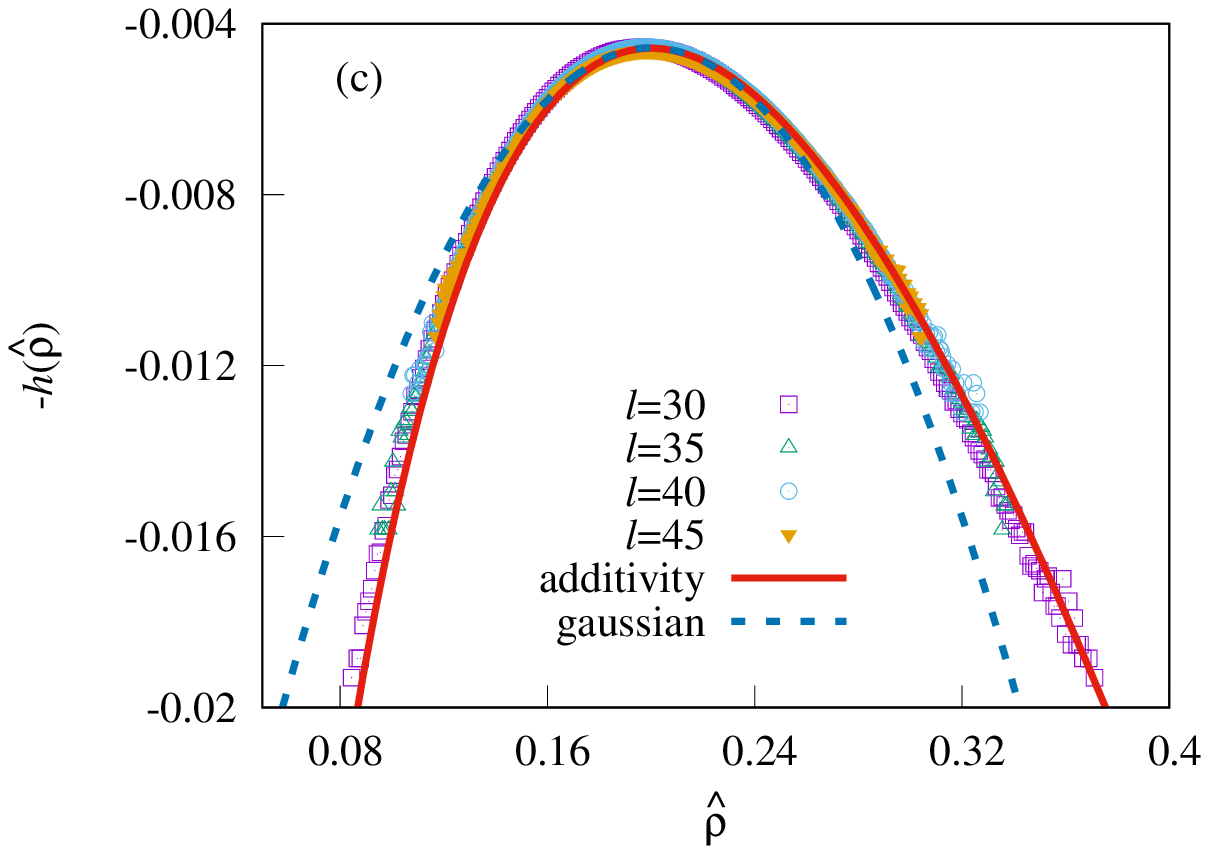}
\caption{{\it Varriant of the Vicsek model.} Large deviation function $-h(\hat \rho)=\ln[{\cal Z}(\mu, v) {\cal P}_v(\hat \rho=n/v)]/v$ [Eq. (\ref{h_rho})] is plotted as a function of the coarse-grained subsystem density $\hat \rho=n/v$, with $n$ and $v$ being subsystem particle-number and volume of the subsystem, respectively, for various subsystem sizes $v=l \times l$ with $l=30$ (violet squares), $35$ (green triangles), $40$ (sky-blue circles), $50$ (yellow inverted triangles) for three sets of noise strength and global density: (i) $\eta=0.5$ and $\rho=0.1$ [panel (a)], (ii) $\eta=0.6$ and $\rho=0.1$ [panel (b)] and (iii) $\eta=0.6$ and $\rho=0.2$ [panel (c)]. Points - simulations, red solid lines - additivity theory, sky-blue dashed lines - corresponding Gaussian large deviation functions. Note that the non-parabolic tails, which are actually due to the non-Gaussian tails in the number distributions, are well captured by additivity theory. }
\label{Prho-VM1}
\end{center}
\end{figure}

\end{widetext}

To calculate subsystem particle-number distribution within additivity, we first calculate in simulations the scaled variance $\sigma^2(\rho)$ of subsystem particle number as a function of density $\rho$. In Fig. \ref{sigma-VM1}, we plot $\sigma^2(\rho)$ as a function of $\rho$ for various values of the noise strengths $\eta=0.4$, $0.5$, $0.6$ and $1.0$. The scaled variance grows quite rapidly in a nonlinear fashion, except at large noise strength where the particles become noninteracting, leading to the linear dependence of the scaled variance on density, i.e., $\sigma^2(\rho) = \rho$ as $\eta \rightarrow \infty$ as in an ideal gas. However, for a finite noise strength, the scaled variance diverges beyond a certain density and the system becomes inhomogeneous with macroscopic particle clusters formed in the system.

Now, using the functional dependence of scaled variance $\sigma^2$ on number density $\rho$, we numerically integrate Eqs. (\ref{mu_rho}) and (\ref{f_rho}) to obtain a nonequilibrium chemical potential and free energy density and to compute the probability distribution $P_v(n)$ of subsystem particle number as given in Eq. (\ref{Pn3}). We also calculate subsystem number distributions from direct microscopic simulations. In Fig. \ref{Pn-VM1}, we plot the number distributions $P_v(n)$ as a function of subsystem particle number $n$ for three different sets of global density and noise strength: (i) $\rho=0.1$ and $\eta=0.6$ (green triangles),  (ii) $\rho=0.1$ and $\eta=0.5$ (violet circles), and (iii) $\rho=0.2$ and $\eta=0.6$ (yellow squares); we take system size $L=500$. We compare the number distributions in simulations with that obtained from additivity theory [Eq. (\ref{Pn3})]. Theory and simulations are in quite good agreement, over several orders of magnitude of probability  $P_v(n)$ and without any fitting parameter involved in our theory.

Next we do a finite size analysis by calculating $\ln[{\cal Z}(\mu, v) {\cal P}_v(\rho)]/v$ of suitably scaled large-deviation probability for different subsystem sizes and check whether large-deviation probabilities indeed converge to that obtained from theory Eq. (\ref{h_rho}). In Fig. \ref{Prho-VM1}, for the above three sets of global density and noise strength - (i), (ii) and (iii), we plot large deviation functions $-h(\hat \rho)=\ln[{\cal Z}(\mu, v) {\cal P}_v(\rho)]/v$ for different subsystem sizes $l=30$ (violet squares), $35$ (green triangles), $40$ (sky-blue circles) and $45$ (yellow inverted triangles), respectively. Then we compare in Fig. \ref{Prho-VM1} the large deviation functions with that computed within additivity (red solid lines) using Eq. (\ref{h_rho}). We find theory and simulations are in quite good agreement with each other. Moreover, in the same figure, we compare the large deviation functions, obtained from theory and simulations, with the ones obtained from the Gaussian distributions with mean subsystem particle number $\langle n \rangle = v \rho$ and variance $\langle n^2 \rangle - \langle n \rangle^2 = v \sigma^2(\rho)$ (sky-blue dashed lines in Fig. \ref{Prho-VM1}). The large deviation functions, beyond the central region around the mean density, deviate significantly from those for the Gaussian ones (logarithm of which are parabolas in Fig. \ref{Prho-VM1}). Indeed, additivity theory captures remarkably well the non-Gaussian (non-parabolic) tails observed in the number distributions (large deviation functions) observed in simulations.

We also attempt a scaling collapse for number fluctuations in the variant of the Vicsek model by rescaling the variance of subsystem particle-number as $\sigma^2(\rho)/\eta^{\Delta}$ and simultaneously rescaling the density $\rho/\eta^{\Delta}$ for various value of noise strengths $\eta=0.4$ (yellow field circles), $0.5$ (red open triangles) and $0.6$ (sky-blue filled triangles); see Fig. \ref{VM-ABP}. In a broad range of density and fluctuation, a reasonably good scaling collapse has been observed for the choice of $\Delta \approx 3.5$. However, we note that, for higher noise strength ($\eta \ge 0.6$) and large density, there are deviations from this particular scaling. Nevertheless, the above rescaled fluctuations provide an insight into comparison between fluctuations with the Vicsek model and its variant.

\subsection{Comparison with other system of self-propelled particles}
\label{comparison}

\begin{figure}[h]
\begin{center}
\leavevmode
\includegraphics[width=9.0cm,angle=0]{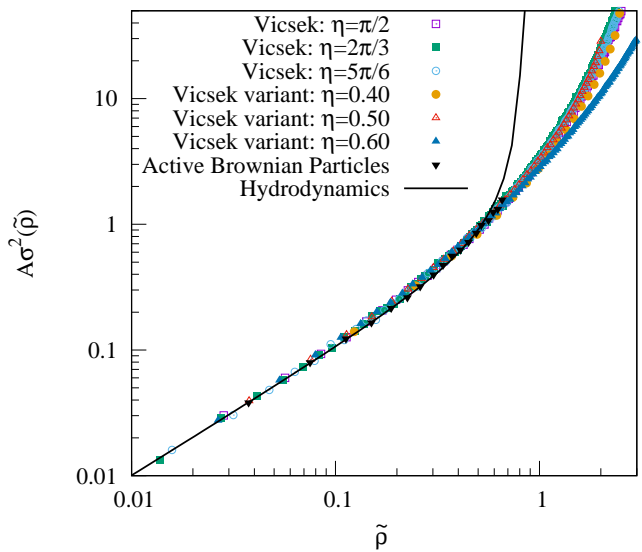}
\caption{Comparison between subsystem density fluctuations in the Vicsek-like models, model of active Brownian particles and a hydrodynamic theory of Ref. \cite{Subhadip_PRE2016}. Rescaled variance of subsystem particle number $A \sigma^2(\tilde \rho)$ is plotted as a function of rescaled subsystem density $\tilde \rho$ for the choice of (i) $A=35/ \eta^{2.5}$ and $\tilde \rho= A  \rho$ in the Vicsek model, (ii) $A=1/ \eta^{3.5}$ and $\tilde \rho = A \rho$ in the variant of the Vicsek model, and (iii) $A=1$ and $\tilde \rho = \rho$ in the model of active Brownian particles. All curves collapse nicely for several orders of magnitudes of the respective variables. We compare the collapsed curves (points) with a hydrodynamimc theory (black line) of Ref. \cite{Subhadip_PRE2016}, which, except in large density regime, compares well with simulations. }
\label{VM-ABP}
\end{center}
\end{figure}

To obtain a broader picture concerning the particle-number fluctuations in the systems of self-propelled particles, in this section we compare the particle-number fluctuations in the Vicsek model, its variant and a system of active Brownian particles \cite{Fily-Marchetti}, which, unlike the Vicsek-like models, does not have any alignment interactions and was previously studied using additivity property \cite{Subhadip_PRE2016}. Indeed, in the disordered isotropic phase, one would expect that, on the  coarse-grained level, the presence of an alignment interaction would possibly change only the relaxation time scale of the polarization density field. In that case, by suitable rescaling of density and fluctuation, it would be possible to have similar characterization of fluctuations, irrespective of the details of the microscopic self-propulsion dynamics. Moreover, the hydrodynamic theory, which was developed in the context of active Brownian particles in Ref. \cite{Subhadip_PRE2016}, could be then expected to capture  fluctuations even in the Vicsek-like models with alignment interactions.

To verify the above assertions, we compare and plot in Fig. \ref{VM-ABP} the scaled subsystem particle-number fluctuations $A \sigma^2(\tilde \rho)$ as a function of the scaled density $\tilde \rho=A \rho$, where the factor $A$ is model-dependent, in the following cases: (i) $A \approx 35/ \eta^{2.5}$ and $\tilde \rho = A  \rho$ in the Vicsek model, (ii) $A \approx 1 \eta^{3.5}$ and $\tilde \rho = A \rho$ in the variant of the Vicsek model and (iii) $A \approx 1$ and $\tilde \rho = \rho$ in the model of active Brownian particles where $\rho$ is the global number density in each case; note that all models are considered in their respective disordered phases. Moreover, in the same Fig. \ref{VM-ABP}, we plot an analytic form of the particle-number fluctuation $\sigma^2(\rho) = [(1+P)\rho(1-\rho/\rho_m)^2]/[1+P(1-\rho/\rho_m)(1-2\rho/\rho_m)]$, derived previously from a hydrodynamic theory [see Eq. (28) in Ref. \cite{Subhadip_PRE2016}], where close-packing density $\rho_m=1.15$ and activity $P=8$ (corresponding to a high Peclet number) are two model-dependent parameters. Interestingly, over several decades of magnitude of density and fluctuation, all curves, except in the variant of the Vicsek model for larger noise strength, collapse on each other quite well. Such a scaling of fluctuations in the self-propelled particle systems with different self-propulsion dynamics suggests a common mechanism through which fluctuations arise in such systems. In the large-density region, especially near phase transitions, there is some disagreement between the hydrodynamic theory and simulations, which is somewhat expected as the linearized hydrodynamic analysis of Ref. \cite{Subhadip_PRE2016} cannot possibly capture the large fluctuations near the transition region.

\section{Summary and concluding remarks}
\label{summary}

In this paper, we study coarse-grained density fluctuations in the disordered phases of the paradigmatic models of self-propelled particles or active matters, which are inherently driven out of equilibrium through a mechanism of self-propulsion of the constituent particles. We consider a broad class of self-propelled particles with alignment interactions, namely the Vicsek models and its variant, consisting of particles, which are point-like, have random self-propulsion velocities, and move in continuum on a two-dimensional periodic space. Particles interact among each other through alignment interactions: At any instant, particles try to follow their neighboring particles in a way so that they align themselves, and move, along a direction, which is obtained by averaging over instantaneous velocities (or directions) of their neighbors. At long times, the systems eventually reach a nonequilibrium steady state.

We coarse-grain the systems by dividing each of the systems into many subsystems and characterize subsystem particle-number (or coarse-grained density) fluctuations through an additivity property, implying a remarkable large-scale thermodynamic structure for the Vicsek-like systems. By using a fluctuation-response relation - the direct consequence of additivity property and employing a numerical scheme developed in this work, we compute the large-deviation probabilities of subsystem particle number, and the corresponding density large-deviation functions in the disordered fluid phase, where macroscopic velocity vanishes in the thermodynamic limit. Importantly, additivity property goes beyond a mean-field analysis because nonequilibrium chemical potential and free energy density function are calculated from the particle-number fluctuation $\sigma^2(\rho)$, which is directly related to spatially integrated density-correlations in the system.
It should be noted that, though there is no macroscopic current in the disordered phase, the systems still remain out-of-equilibrium, due to the violation of detailed balance in the microscopic configuration space, and consequently their microscopic probability weights cannot be described by the equilibrium Boltzmann-Gibbs distribution.

We then compare the density large deviation functions, computed within additivity  theory, with those obtained from direct microscopic  simulations of the models; theoretical and simulation results are found to be in excellent agreement over several orders of magnitude of the large-deviation probabilities of coarse-grained density, without any fitting parameters involved in the theory. Notably, the subsystem particle-number distributions (equivalently, the density large-deviation functions) have non-Gaussian (non-parabolic) tails, which are quite well captured by additivity theory. Our results strongly suggest that the coarse-grained density fluctuations in the self-propelled-particle systems with Vicsek-like interactions are governed by a nonequilibrium chemical potential and a corresponding free energy function. Moreover, comparison between the fluctuations in the disordered phase of several self-propelled particle systems indicates a common mechanism, which gives rise to fluctuations in such systems.

There are a few remarks in order.
In equilibrium systems with the Hamiltonian having short-ranged interactions, the Boltzmann-Gibbs probability weights of microscopic configurations immediately imply additivity \cite{Domb-Lebowitz}. On the other hand, not much is known about the microscopic structure of systems having a nonequilibrium steady state, which cannot be associated with a Hamiltonian as such.      
Interestingly, the Boltzmann-Gibbs distribution, though sufficient, is not necessary for additivity to hold \cite{Eyink1996, Bertin_PRL2006, Chatterjee_PRL2014}. However, the problem of determining the precise conditions, for which additivity would hold in nonequilibrium and a steady-state thermodynamics could be constructed, is a nontrivial one \cite{Bertini_PRL2001, Derrida} and is not yet fully understood \cite{Dickman}. In this scenario, it would be worth checking case-by-case whether a certain class of nonequilibrium systems could possess an additivity property. 
In fact, in most cases, obtaining an analytic expression of the variance of subsystem mass (particle-number) as a function of mass (number) density in an interacting-particle system is difficult and therefore the task of verifying additivity in nonequilibrium systems is also not easy. 
We provide here a simple computational scheme, which could help one to compute density large-deviation functions, and thus to test additivity, in a driven system. 
We emphasize that, although the functional form of the subsystem weight factor $W_v(n)$ in Eq. (\ref{additivity1}) is same throughout the system, additivity property captures variations in coarse-grained density profile $\hat \rho({\bf x})$ through nonequilibrium free energy density functional $f[\hat \rho({\bf x})]$ where ${\bf x}$ is a suitably defined coarse-grained position. More specifically, the postulated additivity property in Eq. (\ref{additivity1}) would immediately imply that the large-deviation probability of an inhomogeneous coarse-grained density profile $\hat \rho({\bf x})$, generated due to spontaneous fluctuations on the subsystem level, can be written as
$$
{\rm Prob.}[\{\hat \rho({\bf x})\}] \sim e^{-V\int d{\bf x} [f(\hat \rho({\bf x}))-f(\rho)-\mu(\rho) (\hat \rho({\bf x}) -\rho)]},
$$ 
where $\rho$ is the global number density and $V$ is the volume of the system. One simplifying aspect in the formulation of additivity is that the large-deviation form of the probability of an inhomogeneous density profile, as in the above equation, does not involve any ``gradient term'' \cite{Cates_PRL2013}, which is analogous to a surface tension between two inhomogeneous phases in an equilibrium system. That is, like in equilibrium, additivity implicitly assumes that the surface tension terms, if any, has only a sub-leading correction to the nonequilibrium free energy functional, which is indeed supported by our results, albeit in the disordered phases.   
In the light of our work, we believe additivity could prove to be a useful concept in characterizing fluctuations not only in systems with Vicsek-like interactions, but also in other active-matter systems.

Moreover, our work brings to the fore some interesting open issues, studies of which could provide further insights into the large-scale properties of self-propelled particles in general. 
Firstly, it remains to be seen whether additivity can be used to characterize properties of Vicsek-like systems near and above criticality, e.g., in the ordered phase, which is known to exhibit giant number fluctuations \cite{review-Marchetti}. 
Secondly, so far we have studied only the static structure of Vicsek-like systems, such as particle-number fluctuations, and have shown that the number fluctuations are related to a nonequilibrium compressibility, through a fluctuation-response relation - reminiscent of the fluctuation-dissipation theorems in equilibrium \cite{Kardar, Green}. However, the interplay between static and dynamic structure, which possibly play an important role in the large-scale behaviors of self-propelled particles, has not been investigated in this work. 
Indeed, unlike in equilibrium \cite{Green}, and except in some simple nonequilibrium processes \cite{Bertini_PRL2001, Sasa, Derrida, Hurtado, Das_PRE2017}, there is no general theoretical understanding of a possible connection between transport and fluctuations in driven many-particle systems. In this context, exact calculations of various transport coefficients can certainly provide some insights into the collective behaviors of self-propelled particles in general. While it may be quite challenging to calculate the transport coefficients in continuum systems, the problem is presumably simpler for systems on a lattice \cite{Kipnis} and is worth pursuing in future.

\section{Acknowledgement}

P.P. thanks Sriram Ramaswamy for useful discussions at the International Centre for Theoretical Sciences (ICTS), Bengaluru, during ``Indian Statistical Physics Community Meeting'' (Code: ICTS/Prog-ISPC/2016/02) and a workshop on ``Stochastic Thermodynamics, Active Matter and Driven Systems'' (Code: ICTS/Prog-STADS/2017/08). S.C. thanks Rakesh Das and Sumanta Kundu for help in computations. We thank two anonymous referees for useful comments and suggestions, and Shradha Mishra for discussions on this work in particular and on the literature on active matters in general. We acknowledge the Thematic Unit of Excellence on Computational Materials Science, funded by the Department of Science and Technology, India for computational facility used in the present study.

\end{document}